# MAGNETOELECTRIC EFFECTS IN
# FERROMAGNETIC/PIEZOELECTRIC MULTILAYER COMPOSITES

G. Srinivasan, C. P. DeVreugd, R. Hayes, M.I. Bichurin* and V.M. Petrov*

*Physics Department, Oakland University, Rochester, MI 4830, USA*
*Novgorod State University, B. S. Peterburgskaya St. 41, 173003*
*Veliky Novgorod, Russia*

## ABSTRACT

*The observation of strong magnetoelectric effects is reported in thick film bilayers and multilayers of ferrite-lead titanate zirconate (PZT) and lanthanum nanganite-PZT. The ferrites used in our studies included pure and zinc substituted cobalt-, nickel- and lithium ferrites. Samples were prepared by sintering 10-40 μm thick films obtained by tape-casting. Measurements of ME voltage coefficients at 10-1000 Hz indicated a giant ME effect in nickel ferrite-PZT, but a relatively weak coupling in other ferrite-PZT and manganite-PZT systems. Multilayers prepared by hot pressing was found to show a higher ME coefficient than sintered samples. Evidence was found for enhancement in ME coefficients when Zn was substituted in ferrites. The Zn-assisted increase was attributed to low anisotropy and high permeability that resulted in favorable magneto-mechanical coupling in the composites. We analyzed the data in terms of our recent comprehensive theory that takes into account actual interface conditions by introducing an interface coupling parameter. Theoretical longitudinal and transverse ME voltage coefficients for unclamped and clamped samples are in general agreement with data. From the analysis we inferred excellent interface coupling for nickel zinc ferrite-PZT and weak coupling for other layered systems.*

## 1. INTRODUCTION

This work is concerned with the fabrication of novel ferromagnetic-ferroelectric thick film multilayers and studies on the nature of magnetoelectric interactions. In such two-phase composites, the magnetoelectric (ME) coupling is mediated by mechanical stress [1]. An applied ac magnetic field produces dynamic deformation in ferromagnets due to magnetostriction and results in an induced electric field due to piezoelectric effect. The systems of interest in the past were bulk samples of nickel or cobalt ferrite with $BaTiO_3$ or lead zirconate titanate (PZT) that showed ME coupling much smaller than predicted values [2-5]. The main cause is low resistivity for ferrites that gives rise to a leakage current and limits the electric field for orienting the dipoles, leading to loss of charges and poor piezoelectric coupling. Such problems could be eliminated in a layered structure [6].

We recently initiated studies on layered heterostructures. The main emphasis of our work has been on ferromagnetic manganite-PZT and ferrite-PZT. Samples were synthesized by sintering thick films made by tape casting and characterized by structural, magnetic, and electrical measurements. Magnetoelectric measurements were made both at low (10 Hz–1 kHz) and high frequencies (9-10 GHz). Key findings and accomplishments are as follows. (i) The first observation of ME coupling in lanthanum strontium manganite-PZT [7]. (ii) A giant low frequency ME interaction in ferrite-PZT [8-11]. (iii) Ultrahigh ME effects at electromechanical resonance (EMR). (iv) Analysis of low frequency ME data using a model for a bilayer that allows the estimation of an all-important interface coupling constant [12,13]. (v) Theoretical models for resonance ME coupling: at EMR for the piezoelectric phase and ferromagnetic resonance for the ferromagnetic phase [14-16]. Our efforts so far have resulted in considerable progress toward an understanding of ME interactions in layered systems [8-19]. The composites are candidate materials for magnetoelectric memory devices, smart sensors, and electric or magnetic field controlled signal-processing devices [20].

In this article, first we provide a brief review of low frequency ME interactions in two-phase product property composites. Section 2 deals with ME effects in bulk composites. In Section 3, our investigations on thick film layered systems and the observation of giant ME effects are described. Theoretical models for bilayers and comparison with data are considered in Section 4.

## 2. MAGNETOELECTRIC EFFECTS IN BULK COMPOSITES

We provide here a review of past efforts by other groups and our current investigations on bulk ferromagnetic-ferroelectric oxides.

The magnetoelectric effect is defined as the dielectric polarization of a material in an applied magnetic field or an induced magnetization in an external electric field [21]. The induced polarization **P** is related to the magnetic field **H** by the expression, **P** = $\alpha$ **H**, where $\alpha$ is the second rank ME-susceptibility tensor. A sample of piezomagnetic-piezoelectric phases is expected to be magnetoelectric since $\alpha = \delta P/\delta H$ is the product of the piezomagnetic deformation $\delta z/\delta H$ and the piezoelectric charge generation $\delta Q/\delta z$ [1]. We are interested in the dynamic ME effect; for an ac magnetic field $\delta H$ applied to a biased sample, one measures the induced voltage $\delta V$. The ME voltage coefficient $\alpha'_E = \delta V/t'\delta H$ and $\alpha = \varepsilon_o \varepsilon_r \alpha'_E$ where t' is the composite thickness and $\varepsilon_r$ is the relative permittivity. The (static) effect, first observed in antiferromagnetic $Cr_2O_3$, is weak in single-phase compounds [22-24].

Bulk composites of interest in the past were $NiFe_2O_4$ (NFO) or $CoFe_2O_4$ (CFO) with $BaTiO_3$ [2-5]. We synthesized similar bulk composites, but with PZT [11,19]. The two oxides were mixed in a ball-mill and disk shaped samples were prepared by traditional sintering at 1400-1500 K or microwave sintering at 1200 K. X-ray diffraction showed no impurities. Magnetic and electrical characterization yielded parameters consistent with expected values for a simple mixed phase. Silver electrodes were deposited on the samples for poling with an electric field perpendicularly to its plane. Samples were placed between the pole pieces of an electromagnet fitted with Helmholtz coils for ME studies and were subjected to a dc field H and ac field (10-1000 Hz) $\delta H=1$ Oe. The resulting voltage $\delta V=t'\delta E$ was measured across the sample thickness. The ME coefficient $\alpha'_E$ was measured for two conditions: (i) transverse or $\alpha'_{E,31}$ for H and $\delta H$ parallel to each other and to the disk plane (*1,2*) and perpendicular to $\delta E$ (direction-*3*) and (ii) longitudinal or $\alpha'_{E,33}$ for all the three fields parallel to each other and perpendicular to sample plane.

Figure 1 show representative data on H dependence of $\alpha'_E$ at 300K and 100 Hz for bulk composites of 50 wt.% ferrite and 50% PZT [11,19,25]. The sample with NFO shows a higher ME voltage than for the composite with CFO. A general increase in $\alpha'_E$ is observed with H to a peak value, followed by a rapid drop. The coefficients are directly proportional to the piezomagnetic coupling q=$\delta\lambda/\delta H$, where $\lambda$ is the magnetostriction, and the H-dependence tracks the slope of $\lambda$ vs H. Saturation of $\lambda$ at high field leads to $\alpha'_E=0$. For most ferrites, $\lambda_{//} = 2 \lambda_\perp$ [19,26] and one expects $\alpha'_{E,33}=2 \alpha'_{E,31}$. The $\alpha'_E$ values are an order of magnitude smaller than theoretical estimates due to poor piezoelectric coupling and leakage current resulting from low composite resistivity [6]. In summary, the low composite resistivity reduces the strength of ME interactions in bulk samples.

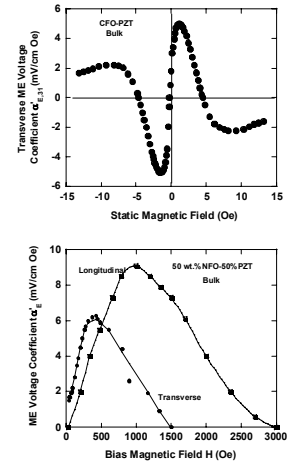

**Fig.1:** *Bias field dependence of ME voltage coefficients in bulk composites of 50 wt.% ferrite and 50% PZT.*

## 3. LAYERED COMPOSITES

The strongest ME coupling is expected in a layered structure due to (i) the absence of leakage current and (ii) ease of poling to align the electric dipoles and strengthen the piezoelectric effect.

Harshe, et al., proposed such structures, provided a theoretical model for a bilayer and prepared multilayers of CFO-PZT or BaTiO$_3$ that showed very weak ME coupling.

Our efforts have been mainly on bilayers and multilayers (MLs) of lanthanum manganites-PZT and ferrite-PZT [8-14]. Layered composites were synthesized using thick films of manganites, ferrites and PZT obtained by tape-casting [27]. The ferrite/manganite powder necessary for tape casting was prepared by the standard ceramic techniques that involved mixing the oxides or carbonates of the constituent metals, followed by presintering and final sintering. A ball-mill was used to grind the powder to submicron size. For PZT films, we used commercially available powder. The fabrication of thick films contained the following main steps: a) preparation of cast of constituent oxides; b) deposition of 10-40 μm thick films tapes by doctor blade techniques; and c) lamination and sintering of composites. Ferrite or PZT powders were mixed with a solvent (ethyl alcohol), plasticizer (butyl benzyl phthalate), and binder (polyvinyl butyral) in a ball mill for 24 hrs. The slurries were cast into 10-40 μm tapes on silicon coated mylar sheets using a tape caster. The films were dried in air for 24 hrs, removed from the mylar substrate and arranged to obtain the desired structure. They were then laminated under high pressure (3000 psi) and high temperature (400 K), and sintered at 1375-1475 K. Bilayers were made with 200 μm thick ferrite or manganite and PZT. Multilayers consisted of (n+1) layers of ferrites/manganite and n layers of PZT (n = 5-30).

Structural characterization was carried out on powdered multilayers and layered samples using an x-ray diffractometer. Two sets of well-defined peaks, corresponding to the magnetic and piezoelectric phases, were present. The data implied a strain-free spinel structure for the ferrites. But the manganite and PZT layers had a strained tetragonal lattice, with a volume reduction of 1-2% [9]. Magnetic characterization included magnetization with a Faraday balance and a vibrating sample magnetometer, ferromagnetic resonance at x-band and magnetostriction with a strain gage. The saturation magnetization agreed with bulk values. Measurements of electrical resistance R and capacitance C were carried out to probe the quality of the composites. The R and C were smaller than expected values due to either higher than expected conductivity of PZT films or the presence of "shorts" in the PZT films since the porosity is on the order of 5-10%. The ferroelectric-to-paraelectric transition temperature of 600 K agreed with expected values [10].

### 3.1. LANTHANUM MANGANITE-PZT

Our studies on manganite-PZT resulted in the first report on ME coupling in the system. Lanthanum manganites with specific concentration of divalent substitutions such as Ba, Ca, or Sr show metallic conduction and ferromagnetism due to the double exchange interactions [28]. The oxides are of interest for studies on ME coupling because of high λ, low resistivity, and structural homogeneity with PZT. Bilayers and multilayers of La$_{0.7}$Sr$_{0.3}$MnO$_3$ (LSMO)-PZT and La$_{0.7}$Ca$_{0.3}$MnO$_3$ (LCMO)-PZT were synthesized. Since the ME voltage is generated exclusively in the piezoelectric phase in layered samples, we measured $\alpha_E$ for unit length of the piezoelectric phase and is related to $\alpha'_E$ through the expression $\alpha_E = \alpha'_E (t'/t)$ where $t'$ and $t$ are the composite and PZT thickness, respectively. Representative results on H-dependence of $\alpha_E$ for manganite-PZT are shown in Fig.2. The data at 120 K is for a LSMO-PZT bilayer. The room temperature value was a factor of 2 smaller than at 120 K [7,10].

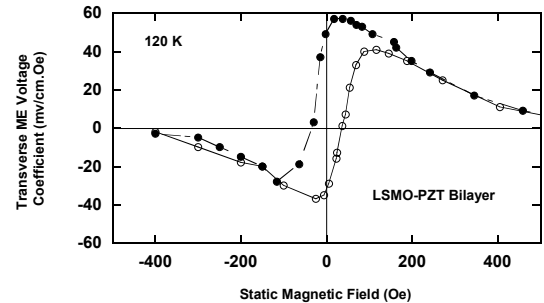

Fig.2: Transverse ME voltage coefficient vs H for lanthanum strontium manganite-PZT bilayer with 200 μm thick manganite and PZT layers.

Strong ME interactions are evident in Fig.2 and one observes unique hysteresis and remanance. The longitudinal coefficient (not shown here) was a factor of 2-3 weaker than the transverse effect. We measured a stronger ME coupling in LSMO-PZT than in LCMO-PZT.

Although XRD shows no impurity phases, the sample parameters were found to be very sensitive to sintering temperature. In addition, the ME effect was weaker in 10-40 micron thick multilayers compared to 200 micron thick bilayers. It further weakens when the layer thickness was reduced or the number of layers was increased. These observations point to the interface diffusion as the possible cause of poor ME effects, a serious problem that needs to be resolved [7,10].

## 3.2. FERRITE-PZT AND GIANT MAGNETOELECTRIC EFFECTS

We succeeded in achieving giant ME effects predicted by theory in bilayers and multilayers of ferrite-PZT. A series of oxides, including pure and Zn substituted nickel-, cobalt, and lithium ferrites was used for the ferromagnetic phase.

### 3.2.1. NICKEL ZINC FERRITE-PZT

We found evidence for a giant ME effect in bilayers and multilayers of nickel zinc ferrite (NZFO), $Ni_{1-x}Zn_xFe_2O_4$ (x=0-0.5), and PZT [8,9]. Representative data on H dependence of $\alpha_E$ are shown in Fig.3 for a multilayer with 15 layers of NFO and 16 layers of PZT. As H is increased, $\alpha_{E,31}$ increases, reaches a maximum value of 400 mV/cm Oe and then drops rapidly to zero. One observes a noticeable hysteresis in the field dependence. The variation of $\alpha_{E,33}$ with H is linear up to 1000 Oe. The longitudinal ME effect is almost an order of magnitude weaker than the transverse effect. We observed a phase difference of 180 degrees between the induced voltages for +H and -H. As discussed later, the magnitude and the field dependence in Fig. 3 are related to variation in q with H. The key observation in Fig.3 is the giant ME coupling that is any order of magnitude stronger than in bulk NFO-PZT (Fig.1). A similar ME coupling, but 10% higher than in multilayers was measured for bilayers of NFO-PZT [8]. The variation in ME coupling with the volume for the two phases was studied in NFO-PZT. Figure 4 shows the measured dependence of $\alpha_E$ on the ratio of volumes v of the magnetostrictivve (m) and piezoelectric (p) phases, $R = {}^m v / {}^p v$, for NFO-PZT. An exponential increase in $\alpha_{E,31}$ occurs with v and shows a maximum of 1500 mV/cm.Oe for R = 2.2. The $\alpha_E$-values in NFO-PZT are one of the

largest ever measured [29-31]. In

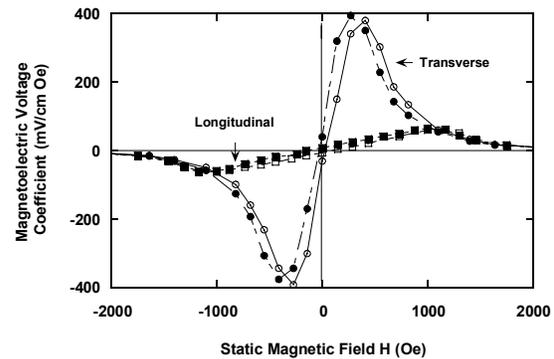

*Fig.3 Magnetoelectric (ME) voltage coefficient $\alpha_E = \delta E / \delta H$ versus bias magnetic field H for a multilayer consisting of 15 layers of NFO and 14 layers of PZT. The data at room temperature and 100 Hz are for transverse (out-of-plane $\delta E$ perpendicular to in-plane $\delta H$) and longitudinal (out-of-plane $\delta E$ and $\delta H$) field orientations. There is 180 deg. phase difference between voltages for +H and –H.*

summary (i) the ME coupling is enhanced in layered samples compared to bulk and (ii) a giant ME effect is evident for NFO-PZT. We recently developed a theoretical model that facilitates quantitative information on interface bonding and is considered in Section 4.

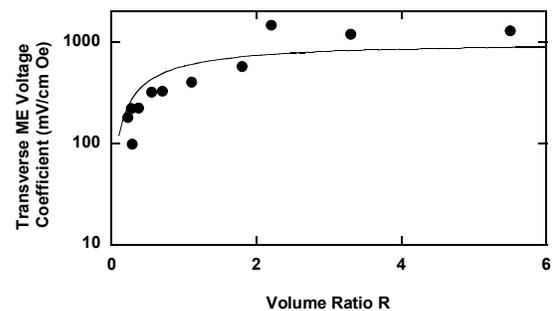

*Fig.4: Variation of the transverse coefficient $\alpha_{E,31}$ with $R = {}^m v / {}^p v$, forNFO-PZT multilayer. The solid line represents theoretical values for a two-layer structure. Values of $\alpha_{E,31}$ are for the bias field H corresponding to maximum value in the ME effect at 100 Hz.*

Similar ME studies were performed on nickel zinc ferrite-PZT samples with x = 0-0.5 [9]. Figure 5

shows representative data on the H-dependence of $\alpha_E$ for NZFO-PZT samples with x = 0-0.4. The data were obtained on samples with n=10-15 at room temperature for a frequency of 100 Hz. For NFO-PZT (x=0), $\alpha_{E,31}$ vs H shows a resonance-like character with a maximum centered at H = 400 Oe. When Zn is substituted for Ni, we notice an increase in the peak value of $\alpha_{E,31}$ for low x-values. As x is increased, a down-shift is observed in H-value corresponding to maximum $\alpha_{E,31}$. The magnetic field range for strong ME effects decreases with increasing Zn content. Data on the longitudinal coupling showed a similar behavior, but the ME coupling is realized over a wide field range. The variations of maximum $\alpha_E$ with x are plotted in Fig.6. Average values of $\alpha_E$ and the spread are shown. The data reveal a 60% increase in the transverse ME voltage coefficient as x is increased from 0 to 0.2, followed by a reduction in $\alpha_{E,31}$ for higher x. The longitudinal coefficient also shows a similar character.

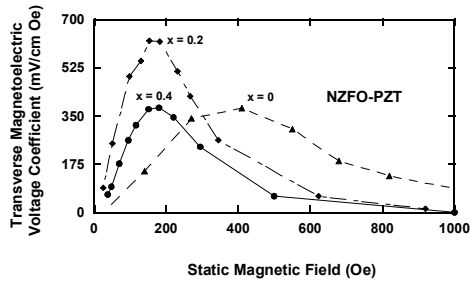

*Fig.5: ME voltage coefficients versus H data for multilayer samples of $Ni_{1-x}Zn_xFe_2O_4$ (NZFO) – PZT with R=1.*

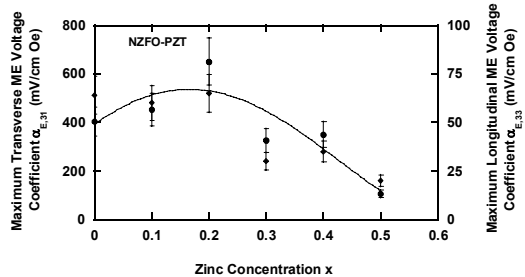

*Fig.6: Zinc concentration dependence of maximum transverse (circles) and longitudinal (squares) ME coefficients in NZFO – PZT layered samples with R=1. The line is guide to the eye.*

Finally, the results discussed so far are for layered samples made by traditional sintering. Since the interface bonding is a critical factor that determines the strength of ME coupling, we prepared similar samples by hot-pressing. The procedure involved sintering at high pressure (5000 psi) and high temperature (1300 K) of preheated multilayers. ME data for such samples are shown in Fig.7 and compared with data for traditionally sintered samples. One observes doubling of peak ME voltage coefficient for hot-pressed case compared to sintered samples. This study is in progress.

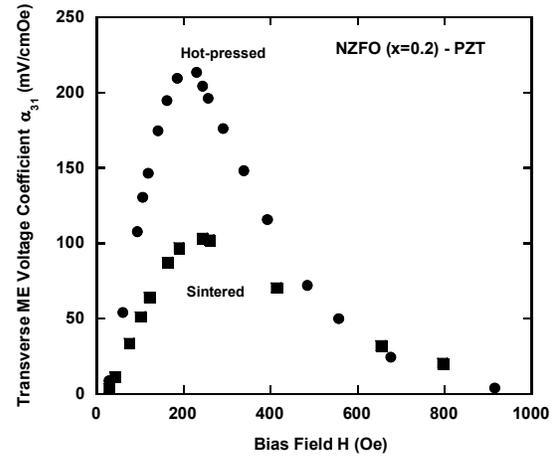

*Fig.7: Low frequency ME voltage coefficient vs H data for hot-pressed and sintered multilayers of NZFO (x-0.2)-PZT.*

Now we comment on the results for NZFO-PZT layered composites. The most significant inferences in Fig.2-6 are: (i) for equal volume of ferrite and PZT, the maximum $\alpha_{E,31}$ ranges from 400 mV/cm Oe in multilayers to 460 mV/cm Oe in the bilayer, (ii) $\alpha_{E,31}$ increases with increasing volume of nickel ferrite and the largest measured value is 1500 mV/cm Oe for $^mv/^pv$ =2.2 and (iii) Zn-substitution in NFO results in the strengthening of ME coupling with the maximum occuring for x=0.2. These ME coefficients must be compared with 20 mV/cm Oe for $Cr_2O_3$, the best single phase ME material [22-24]. It is more than an order of magnitude higher than reported values for ferrite – $BaTiO_3$ *bulk* composites, and a factor of five larger than in laminated composites of Ni(Co,Cu)-Mn ferrite – PZT [5,30].

### 3.2.2.    COBALT ZINC FERRITE – PZT

Next we consider studies on samples of CZFO-PZT [9,25].  Figure 8 shows representative data on the H dependence of $\alpha_{E,31}$ and $\alpha_{E,33}$ for a multilayer sample in which 40% Co is replaced by Zn in cobalt ferrite.    The data at room temperature and 100 Hz are for a sample with n = 10.  As the bias field is increased from zero, $\alpha_E$ increases rapidly to a peak value.  With further increase in H, the ME coefficients drop to a minimum or zero value.  There was no noticeable hysteresis or remanence in $\alpha_E$ vs. H.  Consider the H dependence of transverse and longitudinal coefficients.  Although overall features in Fig.7 are similar for both cases, one finds the following differences.   (i) The initial rate of increase in $\alpha_E$ with H is much higher for         the transverse case than for longitudinal orientation for the fields.   (ii) The peak-$\alpha_{E,31}$ is a factor of five higher than $\alpha_{E,33}$. (ii) The peak value in $\alpha_{E,33}$ occurs for a higher bias field than for the transverse case.  These observations could be understood in terms of H variation of parallel and perpendicular magnetostriction for the ferrite.

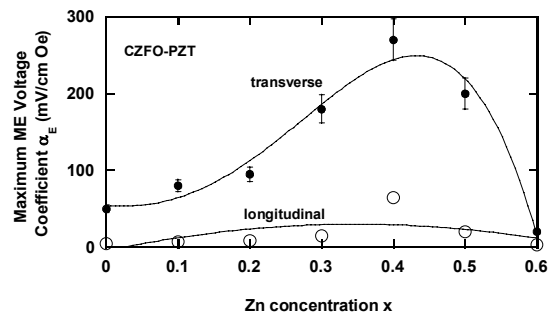

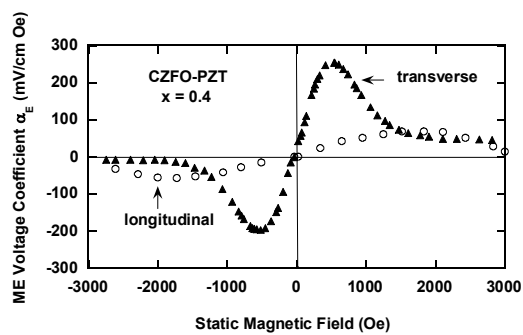

*Fig.8:Static field dependence of room temperature transverse and longitudinal ME voltage coefficients, $\alpha_{E,31}$ and $\alpha_{E,33}$ respectively, for a multilayer composite of $Co_{1-x}Zn_xFe_2O_4$ (CZFO) (x=0.4)–lead zirconate titanate (PZT). The sample contained 11 layers of CZFO and 10 layers of PZT with a thickness of 18 μm.*

Similar $\alpha_E$ vs H data were obtained for samples with x-values varying from 0 to 0.6.  Both $\alpha_{E,31}$ and $\alpha_{E,33}$ were measured. Figure 9 shows the variation of maximum $\alpha_E$ with x for both transverse and longitudinal cases. The peak coefficients were

*Fig.9:  Variation of peak transverse and longitudinal ME voltage coefficients with zinc concentration x in layered CZFO – PZT.  The data at room temperature and 100 Hz were obtained from profiles as in Fig.8.  The bars indicate the range of measured values.  The solid lines are guide to the eyes.*

measured on several samples and the figure shows the range of measured values and their average.  As the Zn
substitution is increased, one observes a sharp increase in $\alpha_{E,31}$, from 50 mV/cm Oe for x = 0 to 280 mV/cm Oe for x = 0.4.   Further increase in x is accompanied by a substantial reduction in $\alpha_{E,31}$. The longitudinal coupling parameter is very weak for the entire series.  One needs to compare the results in Figs.8-9 with past studies on bulk samples of CFO-BaTiO3 and CFO-PZT and multilayers of CFO-PZT.[16]  Bulk samples showed very weak ME interactions (see Fig.1) but layered CFO-PZT showed $\alpha_E$ of 75 mV/cm Oe, comparable to values in Fig.9 [6].  It is obvious from the present study that Zn substitution in cobalt ferrite is a key ingredient for strong ME coupling in multilayers [9]. We attribute the efficient field conversion properties to modification of magnetic parameters due to Zn (Section 4).

### 3.2.3. LITHIUM ZINC FERRITE - PZT

Finally, we consider studies on lithium zinc ferrite-PZT composites [32]. Samples with n=10 and 15, a layer thickness of 15 micron and ferrites of composition $Li_{0.5-x/2}Zn_xFe_{2.5-x/2}O_4$ for x=0-0.4 were synthesized. Figure 10 shows data on H-variation of the transverse ME voltage coefficient for x = 0-0.3. The data are for a frequency of 100 Hz at room temperature. To our knowledge, these data constitute the first report of strong ME coupling in lithium ferrite – PZT composites of any kind. Important observations are as follows. (i) Data show features similar to the other two systems. (ii) A factor of five increase in the peak $\alpha_{E,31}$ value is evident when x is increased from 0 to 0.3. (iii) An up-shift in the H-value corresponding the peak $\alpha_{E,31}$ is seen as x is increased from 0. Recall that for CZFO-PZT and NZFO-PZT samples, a down-shift in H for maximum $\alpha_{E,31}$ is observed for increasing x. (iv) The H-interval for strong ME effects is essentially independent of x. The inset in Fig.10 shows data on peak values of $\alpha_{E,31}$ vs x. One notices a rapid increase in the ME voltage coefficient with x for x = 0- 0.3, followed by a sharp decrease for x=0.4 [32].

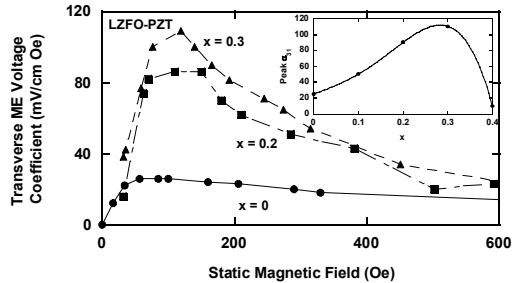

*Fig. 10: Transverse voltage coefficient versus H profiles for multilayer samples of $Li_{0.5-x/2}Zn_xFe_{2.5-x/2}O_4$ (LZFO) - PZT. The inset shows peak value of $\alpha_{E,31}$ vs x.*

## 4. Theory

We developed a model for ME effects in layered samples for an understanding of the giant ME effects [12,13]. Previous attempts just considered longitudinal ME voltage coefficient $\alpha_{E,33}$ for ideal coupling at the interface [6]. The major deficiencies of the earlier model are as follows. (i) For the longitudinal case, influence of the finite magnetic permeability for the ferrite was ignored. A reduction in the internal magnetic field and weakening of ME interactions are expected due to demagnetizing fields. (ii) The model did not consider ME coupling under transverse field orientations for which studies on ferrite-PZT show a giant ME effect. (iii) It is necessary to quantify less-than-ideal interface coupling [6]. We present here a comprehensive theory in which the composite is considered as a homogeneous medium with piezoelectric and magnetostrictive subsystems. Important aspects of the model are as follows. (i) We take into account less-than-ideal interface conditions by introducing an interface coupling parameter *k*. (ii) Expressions for longitudinal and transverse ME coefficients are obtained using the solutions of elastostatic and electrostatic equations. (iii) We consider a third field orientation of importance in which $E$, $\delta E$, $H$, and $\delta H$ are in the sample plane and parallel to each other and is referred to as in-plane longitudinal ME coupling ($\alpha_{E,11}$). (iv) The theory is developed for two types of measurement conditions: unclamped and clamped bilayers. (v) The ME voltage coefficients are estimated from known material parameters (piezoelectric coupling, magnetostriction, elastic constants, etc.,) and are compared with data.

We consider a bilayer in the *(1,2)*-plane consisting of piezoelectric and magnetostrictive phases as shown in Fig.11. Since most studies have dealt with polycrystalline thick films for the two constituents, we do not assume any epitaxial characteristics for the layers. An averaging method is used for deriving effective composite parameters and is carried out in two stages.[17-19] In the first stage, the sample is considered as a bilayer. In the second stage, the bilayer is considered as a homogeneous medium.

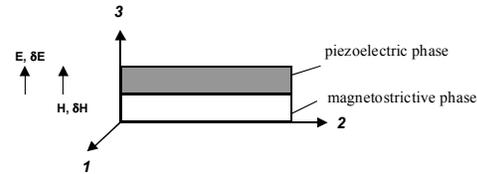

*Fig.11: Schematic diagram showing a bilayer of piezoelectric and magnetostrictive phases in the (1,2) plane. Field orientations for longitudinal magnetoelectric (ME) effect are also shown.*

The theory yields the following expressions for the longitudinal and transverse voltage coefficients for unclamped bilayers [12,13].

$$\alpha_{E,33} = \frac{-2\mu_0 k(1-v)\,^p d_{13}\,^m q_{13}}{2\,(^p d_{13})^2(1-v)k + {}^p\varepsilon^T_{33}\,[(^p s_{11} + {}^p s_{12})(v-1) - k(^m s_{11} + {}^m s_{12})v]} \times$$

$$\times \frac{[(^p s_{11} + {}^p s_{12})(v-1) - (^m s_{11} + {}^m s_{12})kv]}{[\mu_0(v-1) - {}^m\mu_{33}v][kv(^m s_{11} + {}^m s_{12}) - (^p s_{11} + {}^p s_{12})(v-1)] + 2\,(^m q_{31})^2 kv^2} \qquad (1)$$

$$\alpha_{E,31} = \frac{-k(1-v)\,^p d_{13}\,(^m q_{11} + {}^m q_{12})}{(^m s_{11} + {}^m s_{12})^p\varepsilon^T_{33}\,kv + (^p s_{11} + {}^p s_{12})\,^p\varepsilon^T_{33}\,(1-v) - 2\,(^p d_{13})^2\,k(1-v)} \qquad (2)$$

Here d and q are the piezoelectric and piezomagnetic coupling coefficients, respectively, s is the compliance coefficient, $\varepsilon^T$ is permittivity at constant stress, $\mu$ is the tensor permeability and $v = {}^p v/(^p v + {}^m v)$. The parameter k is the interface coupling parameter, with k=1 for ideal coupling and k=0 for the case with no friction.

We now use the model for theoretical estimates of $\alpha_E$ for comparison with the data. Calculated $\alpha_{E,33}$ are expected to be quite small due to weak $q_{13}$ and demagnetizing fields. Such a prediction is in agreement with observations in all of our systems discussed in Sec.3. Discussion to follow is therefore restricted to transverse ME coupling. We estimated $\alpha_{E,31}$ vs H for manganite-PZT and ferrite-PZT using bulk values for s, $\varepsilon$ and d, and q=$q_{11}$+$q_{12}$ obtained from $\lambda$ vs H data as in Fig. 12. The following values were used in Eq. (2) for the composite parameters:

$^p s_{11} = 15*10^{-12}$ m$^2$/N, $^p s_{12} = -5*10^{-12}$ m$^2$/N, $^m s_{11} = 6.5*10^{-12}$ m$^2$/N; $^m s_{12} = -2.4*10^{-12}$ m$^2$/N, and $\varepsilon_{33}/\varepsilon_0 = 1750$ [6,9,33]. The measured value for d$_{33}$ in the bulk and layered samples was 250 pm/V, corresponding to d$_{31}$ = 0.5 d$_{33}$ = 125 pm/V.

Theoretical estimates of ME coupling for NZFO – PZT indicated very good agreement with data for the entire series of Zn substitution [9]. Representative results are shown in Fig.13 for x=0. The estimates are for a series of interface coupling k and q-values obtained from magnetostriction data in Fig.12 and other material parameters mentioned earlier for ferrites. The data were obtained on a multilayers with n=15 and a layer thickness of 15 $\mu$m. One observes good agreement between theory for k=1 and data. There is excellent agreement in the magnitude of $\alpha_{E,31}$, but the theory predicts a sharper drop in $\alpha_{E,31}$ at high fields than observed experimentally.

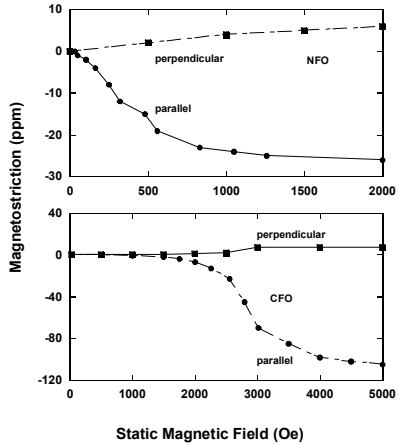

Fig.12: Room temperature in-plane parallel ($\lambda_{11}$) and perpendicular ($\lambda_{12}$) magnetostriction versus H for NFO and CFO bulk samples made from thick films.

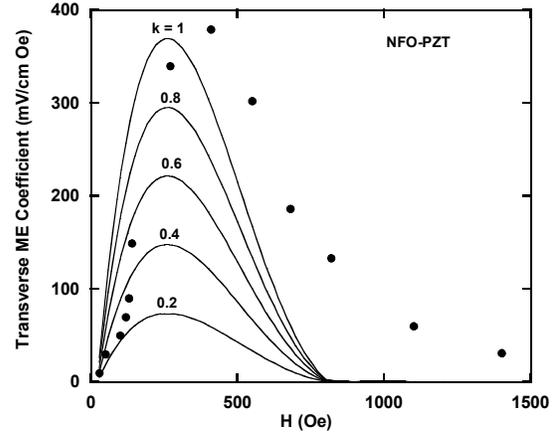

Fig.13: Theoretical (lines) and measured (circles) $\alpha_{E,31}$ versus H for NFO – PZT composites.

A similar comparison of the data and theoretical values of $\alpha_{E,31}$ for CZFO-PZT, however, revealed a poor interface coupling [9].

Representative results are shown in Fig.14 for x = 0. We observe a substantial disagreement between theory and data. Neither the magnitude of $\alpha_{E,31}$ for k=1 nor its H dependence agree with the data. The predicted values for perfect interface coupling are an order of magnitude higher and H-values for maximum $\alpha_{E,31}$ are a lot smaller than measured values. A weak interface coupling, with k on the order of 0.1, is evident for x=0. We noticed a similar behavior for x=0.2. Magnitudes of theoretical $\alpha_{E,31}$ for k=0.2 were are in agreement with the data and was indicative of a stronger interface coupling than in x=0. A general improvement in the ME coupling is thus accomplished with Zn substitution in the ferrite.

It is appropriate to compare the results of theory for the two ferrite-PZT systems. In order to facilitate such a comparison, the interface coupling obtained from theory are shown in Fig.15. One draws the following inferences from Fig.13-15. First, for CZFO–PZT, there is total lack of agreement between theory for k=1 and data. Although magnetostriction data implies strong piezomagnetic coupling, the estimated $\alpha_{E,31}$ are factor of 2-10 higher than measured values. Second, the introduction of Zn leads to an enhancement in the strength of interface coupling k and $\alpha_E$, in particular for CZFO-PZT. The constant k increases from 0.1 to 0.6 when Zn progressively replaces 40% of Co. A near perfect coupling for NZFO-PZT is inferred from the calculations. Possible causes of the wide variations in k are discussed in the following section. The key accomplishment here is the ability to obtain information on interface coupling [9].

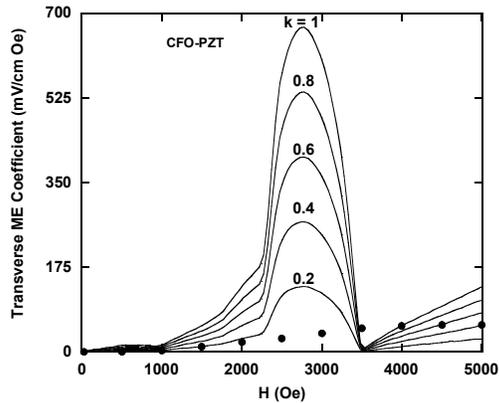

*Fig.14:* *Comparison of theoretical and measured values of the transverse ME voltage coefficient $\alpha_{E,31}$ for layered samples of CFO– PZT. The solid curves are theoretical values for a series for interface coupling parameter k.*

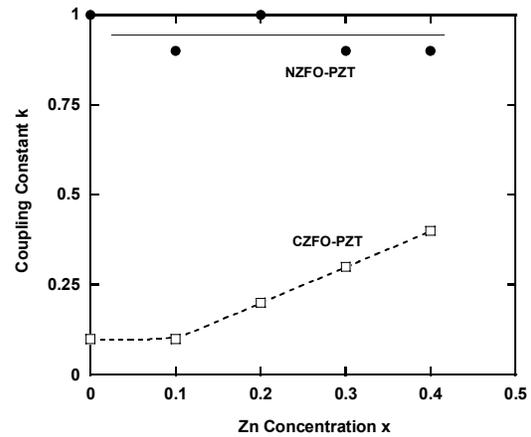

*Fig. 15: Interface coupling k as a function of Zn substitution x in ferrites.*

Analysis of the low frequency ME data was also performed for LZFO-PZT composites [32]. A weak interface coupling with k = 0.2 was obtained for LFO-PZT, and k increased with Zn substitution in the ferrite.

A similar theoretical analysis was carried out for LSMO-PZT and the results are shown in Fig.16. The magnetostriction data needed for q-values is also shown. The calculated values for k=1 are an order of magnitude higher than the data [7].

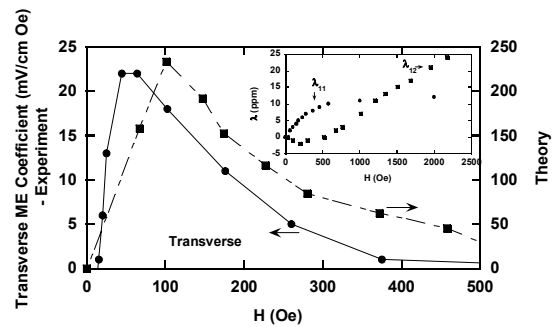

*Fig.16: Comparison of theoretical estimates and data for transverse ME voltage coefficient in LSMO-PZT bilayers.*

In summary, the theory and analysis presented here provide an elegant means to quantify an otherwise complex parameter k. The analysis reveal excellent interface coupling only for NZFO-PZT layered systems. We need to point out an important limitation of the model used here. The model is valid for a simple bilayer structure. It is necessary to extend the theory to include a multilayer consisting of n-interfaces in a structure with (n+1)-ferrites and n-PZT layers. One expects a stronger interface coupling in multilayers than in bilayers.

## 4.1 ORIGIN OF GIANT ME EFFECTS

Next we comment on the possible causes of giant ME effects and the inferred excellent coupling parameter k only in NZFO-PZT. One expects k to be dependent on a variety of factors including structural, mechanical, chemical and electromagnetic parameters. X-ray diffraction data indicated the absence of any structural abnormalities. Mechanical bonding at the interface that arises due to the high temperature processing of the composite is another key factor that determines k. In some recent works, the bonding between magnetostrictive and piezoelectric media was accomplished with the use of silver epoxy [30]. The understanding of interface coupling in such cases is further complicated by the introduction of a foreign material. In our sintered composites it is difficult to quantify the strength of mechanical bonding. Simple peel tests are useful for such information, for example, in metal-polymer samples. We are not aware of such tests for sintered composites. One also expects k to be influenced by the possible presence of chemical inhomogeneities at the interface. An observation of importance in this regard is the deterioration in ME coupling in CZFO-PZT samples sintered at the high end of the temperature range 1375-1475 K. Although x-ray diffraction data did not indicate any detectable levels of impurities, a "fused" interface and sample warping indicate microscopic chemical inhomogeneities in CZFO-PZT that will have adverse impact on k.

Finally we address the anticipated effects of electromagnetic parameters on k. For answers, one needs to focus on the effects of magnetostriction and magnetoelastic coupling for the following reasons. PZT is the only piezoelectric phase used in all the composites and its lattice is strained in both systems. The selection of composites and measurements are directed toward examining the role of ferrites/manganites on (i) ME voltage and (ii) the dependence of $\alpha_{E,31}$ on H. There are two types of magnetostriction in a ferromagnet: (i) Joule magnetostriction associated with domain movements and (ii) volume magnetostriction associated with magnetic phase change. The volume magnetostriction is not important in the present situation since it is significant only at temperatures close to the Curie temperature. In ferrites domains are spontaneously deformed in the magnetization direction. Under the influence of a bias field H and ac field $\delta H$, domain wall motion and domain rotation contribute to the Joule magnetostriction. Since the ME coupling involves dynamic magneto-mechanical coupling, key requirements for the ferrite are unimpeded domain wall motion, domain rotation and a large $\lambda$. A soft, high initial permeability (and low anisotropy as indicated by FMR studies) ferrite, such as NFO, is the key ingredient for strong ME effects. In magnetically hard cobalt ferrite, however, one has the disadvantage of a large anisotropy field that limits domain rotation. Our magnetization measurements yielded an initial permeability of 20 for NFO vs 3.5 for CFO. Figure 17 shows the composition dependence of the initial permeability $\mu_i$ for CZFO and NZFO [33]. With increasing Zn concentration in CZFO-PZT, $\mu_i$ increases by a factor of 30 to a maximum value for x = 0.5 in agreement with anticipated reduction in the anisotropy. For NZFO, $\mu_i$ shows an order of magnitude increase for the composition range of importance. Even though $\mu_i$ and $\alpha_{E,31}$ tracks each other in CZFO-PZT, a similar behavior is absent in NZFO-PZT. It is clear that the important factor is not just $\mu_i$ but the magneto-mechanical coupling $k_m$ given by $k_m = (4\pi\lambda'\mu_r/E)^{1/2}$ where $\lambda'$ is the dynamic magnetostrictive constant and $\mu_r$ is the reversible permeability, parameters analogous to $q=(q_{11}+q_{12})$ and $\mu_i$, respectively, and E is the Young's modulus [34]. In Fig.18, we compare maximum values of $\alpha_{E,31}$ and the product $\mu_i q$ as a function of x for both systems. The factors rise and fall in tandem and the maxima occur around the same composition, i.e., x = 0.2-0.3 for NZFO-PZT and 0.3-0.4 for CZFO- PZT. Thus it is logical to associate the strong interface coupling for the entire series of NZFO-PZT and Zn-rich CZFO-PZT with efficient magneto-mechanical coupling.

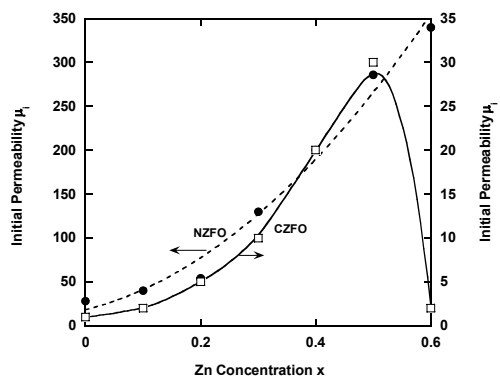

*Fig.17: Composition dependence of the initial permeability $\mu_i$ (from Ref.33) for CZFO and NZFO. The lines are guide to the eye.*

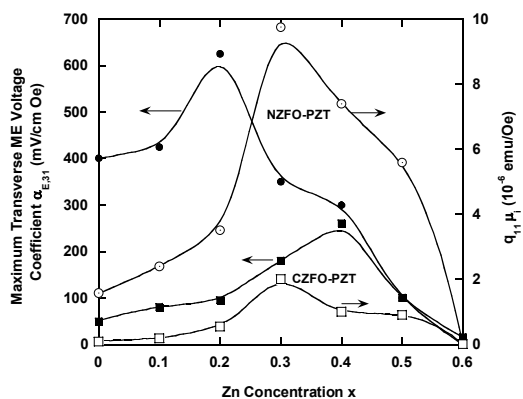

Fig.18: *Comparison of Zn substitution dependence of maximum values of $\alpha_{E,31}$ and the product $q\mu_i$ for NZFO-PZT and CZFO-PZT. The parameter $q = q_{11}+q_{12}$ is the net piezomagnetic coupling constant estimated from the slope of $\lambda$ vs H data.*

## 5. CONCLUSIONS

Studies on layered samples of lanthanum manganite-PZT and pure and zinc substituted ferrites - PZT show evidence for strong ME interactions. Of particular interest is the giant ME voltage coefficients in nickel ferrite-PZT. The voltage coefficient $\alpha_E$ show an overall increase with increasing Zn concentration x in CZFO-PZT and NZFO-PZT. A maximum in $\alpha_E$ occurs for x = 0.2-0.4, depending on the ferrite. Analysis of the data using our model for a bilayer reveals ideal interface conditions for NZFO-PZT. The data implies poor coupling in other ferrite-PZT and manganite-PZT. The Zn assisted enhancement in ME coefficient is primarily due to low anisotropy and high permeability for the ferrites that result in favorable magneto-mechanical coupling in the composites.